%% file: main.tex
\documentclass[10pt,conference]{IEEEtran}
\IEEEoverridecommandlockouts

\usepackage{cite}
\usepackage{amsmath,amssymb,amsfonts}
\usepackage{algorithmic}
\usepackage{textcomp}
\usepackage{xcolor}
\usepackage{xspace}
\usepackage{pifont}%
\usepackage[inkscapelatex=false]{svg}
\usepackage{booktabs}
\usepackage{listings}
\usepackage{graphicx,xfp,subcaption}
\usepackage{tabularray}
\usepackage{tabularx}
\usepackage{tcolorbox}
\usepackage{multirow}
\usepackage{caption,array}
\usepackage{url}
\usepackage{hyperref}
\usepackage{array}
\usepackage{balance}
\usepackage[normalem]{ulem}
\usepackage{pgfplots, pgfplotstable}
\pgfplotsset{compat=1.18}

\def\BibTeX{{\rm B\kern-.05em{\sc i\kern-.025em b}\kern-.08em
    T\kern-.1667em\lower.7ex\hbox{E}\kern-.125emX}}

\input{figures/data}
\begin{document}

\NewDocumentCommand{\codeword}{v}{%
\small{\texttt{{#1}}}%
}

\newcommand{\ie}{\emph{i.e.,}\xspace}
\newcommand{\eg}{\emph{e.g.,}\xspace}
\newcommand{\cmark}{\ding{51}}%
\newcommand{\xmark}{\ding{55}}

\input{macros}

\newcommand{\gulzar}[1]{\textcolor{red}{G: #1}}
\newcommand{\tien}[1]{\textcolor{blue}{T: #1}}
\newcommand{\question}[1]{\textcolor{brown}{Question: #1}}
\newcommand{\waris}[1]{\textcolor{orange}{W: #1}}
\newcommand{\edit}[1]{\textcolor{cyan}{#1}}

\newcommand{\todo}[1]{\textcolor{green}{TODO: #1}}

\title{Are the Majority of Public Computational Notebooks Pathologically Non-Executable?}

\author{\IEEEauthorblockN{Tien Nguyen, Waris Gill, Muhammad Ali Gulzar}
\IEEEauthorblockA{Virginia Tech\\
Blacksburg, USA \\
\{tiennguyen, waris, gulzar\}@vt.edu}
}

\maketitle
\input{sections/abstract}

\input{sections/s1_intro}

\input{sections/s2_motivation}

\input{sections/s3_research_questions}

\input{sections/s4_approach}

\input{sections/s5_results}
\input{sections/s6_discussion}
\input{sections/s7_related_work}

\input{sections/s8_conclusion}

\balance

\input{main.bbl}
\end{document}

%% file: macros.tex
\def\totalRepos{4,456\xspace}
\def\totalNotebooksInDataset{42,546\xspace}
\def\totalTimeout{1,102\xspace}
\def\percentTimeout{2.5\%\xspace}
\def\totalExecutable{7,887\xspace}
\def\percentExecutable{18.5\%\xspace}
\def\totalNonExecutable{34,659\xspace}
\def\percentNonExecutable{81.5\%\xspace}
\def\totalPathological{7,387\xspace}
\def\percentPathological{21.3\%\xspace}
\def\averagePercentPartialPathological{34.1\%\xspace}
\def\totalPathologicalOverFiftyPercent{1,270\xspace}
\def\percentPathologicalOverFiftyPercent{17.2\%\xspace}
\def\totalPathologicalZeroPercent{2,059\xspace}
\def\percentPathologicalZeroPercent{27.9\%\xspace}
\def\overallExecutableCells{31.9\%\xspace}
\def\totalRestorable{27,272\xspace}
\def\percentRestorableInNonExecutable{78.7\%\xspace}
\def\fullyRestored{1,463\xspace}
\def\percentFullyRestoredBefore{16.1\%\xspace}
\def\percentFullyRestored{5.4\%\xspace}
\def\averageIncreaseFullyRestored{84.1\%\xspace}
\def\partiallyRestored{3,480\xspace}
\def\percentPartiallyRestored{12.8\%\xspace}
\def\percentPartiallyRestoredBefore{5\%\xspace}
\def\percentPartiallyRestoredAfter{31.1\%\xspace}
\def\averageIncreasePartiallyRestored{26.1\%\xspace}
\def\totalRestored{4,943\xspace}
\def\percentRestored{18.1\%\xspace}
\def\failedToRestore{22,329\xspace}
\def\percentFailedToRestore{81.9\%\xspace}
\def\totalFileNotFound{4,546\xspace}
\def\percentFileNotFoundInRestorable{16.7\%\xspace}
\def\percentFileNotFoundInNonExecutable{13.1\%\xspace}
\def\percentFileNotFoundInTotal{10.7\%\xspace}
\def\totalFileNotFoundRestored{1,378\xspace}
\def\percentFileNotFoundRestored{30.3\%\xspace}
\def\totalFileNotFoundNotRestored{3,168\xspace}
\def\percentFileNotFoundNotRestored{69.7\%\xspace}
\def\totalFailedToGenerateFiles{817\xspace}
\def\percentFailedToGenerateFiles{18\%\xspace}
\def\totalSuccessfulToGenerateFiles{3,729\xspace}
\def\percentSuccessfulToGenerateFiles{82\%\xspace}
\def\percentFileNotFoundFullyRestored{4.8\%\xspace}
\def\percentFileNotFoundFullyRestoredInSuccessfullyGenerated{5.9\%\xspace}
\def\percentFileNotFoundPartiallyRestoredInSuccessfullyGenerated{31.1\%\xspace}
\def\percentFileNotFoundPartiallyRestored{25.5\%\xspace}
\def\percentFileNotFoundRestoredInSuccessfullyGenerated{37\%\xspace}
\def\avgIncreaseAfterFileFixed{28\%\xspace}
\def\totalFileNotFoundLessFivePercent{148\xspace}
\def\totalModuleNotFound{23,476\xspace}
\def\percentModuleNotFound{55.2\%\xspace}
\def\percentModuleNotFoundInTotal{55.2\%\xspace}
\def\percentModuleNotFoundInRestorable{86.1\%\xspace}
\def\percentModuleNotFoundInNonExecutable{67.7\%\xspace}
\def\totalInvalidModuleNotFound{5,970\xspace}
\def\percentInvalidModuleNotFoundInAllModuleNotFound{25.4\%\xspace}
\def\totalValidModuleNotFound{17,506\xspace}
\def\percentValidModuleNotFoundInAllModuleNotFound{74.6\%\xspace}
\def\percentValidModuleNotFoundInNonExecutable{50.5\%\xspace}
\def\totalModuleNotFoundRestored{3,760\xspace}
\def\percentModuleNotFoundRestored{21.5\%\xspace}
\def\totalModuleNotFoundNotRestored{13,746\xspace}
\def\percentModuleNotFoundNotRestored{78.5\%\xspace}
\def\percentModuleNotFoundFullyRestored{6.8\%\xspace}
\def\percentModuleNotFoundFullyRestoredInTotalFixed{31.5\%\xspace}
\def\percentModuleNotFoundPartiallyRestored{14.7\%\xspace}
\def\percentModuleNotFoundPartiallyRestoredInTotalFixed{68.5\%\xspace}
\def\averagePercentModuleNotFoundRestoredIncrease{40.5\%\xspace}
\def\totalNameError{404\xspace}
\def\percentNameErrorInTotal{0.9\%\xspace}
\def\percentNameErrorInRestorable{1.5\%\xspace}
\def\percentNameErrorInNonExecutable{1.2\%\xspace}
\def\totalUndefined{363\xspace}
\def\percentUndefinedInNameError{89.9\%\xspace}
\def\totalDefinedAfter{5\xspace}
\def\percentDefinedAfterInNameError{1.2\%\xspace}
\def\totalBothDefinedAndUndefined{36\xspace}
\def\percentBothDefinedAndUndefinedInNameError{8.9\%\xspace}
\def\totalNameErrorFixed{183\xspace}
\def\percentNameErrorFixedInAllNameErrors{45.3\%\xspace}
\def\percentNameErrorFixedInNonExecutable{0.5\%\xspace}
\def\totalNameErrorFixedFull{35\xspace}
\def\percentNameErrorFixedFullInTotalFixed{19.1\%\xspace}
\def\percentNameErrorFixedFullInAllNameErrors{8.7\%\xspace}
\def\totalNameErrorFixedPartial{148\xspace}
\def\percentNameErrorFixedPartialInTotalFixed{80.9\%\xspace}
\def\percentNameErrorFixedPartialInAllNameErrors{36.6\%\xspace}
\def\avgIncreaseAfterNameFixed{9.4\%\xspace}
\def\totalRequirement{1,150\xspace}
\def\percentRequirementInTotal{25.8\%\xspace}
\def\totalUserInputs{469\xspace}
\def\percentUserInputsInNonExecutable{1.4\%\xspace}
\def\percentUserInputsInTotal{1.1\%\xspace}
\def\totalAttributeError{1,917\xspace}
\def\percentAttributeErrorInNonExecutable{5.5\%\xspace}
\def\percentAttributeErrorInTotal{4.5\%\xspace}
\def\totalImportError{1,723\xspace}
\def\percentImportErrorInNonExecutable{5\%\xspace}
\def\percentImportErrorInTotal{4\%\xspace}
\def\totalValueError{1,705\xspace}
\def\percentValueErrorInNonExecutable{4.9\%\xspace}
\def\percentValueErrorInTotal{4\%\xspace}
\def\totalTypeError{1,505\xspace}
\def\percentTypeErrorInNonExecutable{4.3\%\xspace}
\def\percentTypeErrorInTotal{3.5\%\xspace}
\def\totalKeyError{1,018\xspace}
\def\percentKeyErrorInNonExecutable{2.9\%\xspace}
\def\percentKeyErrorInTotal{2.4\%\xspace}
\def\totalIndexError{446\xspace}
\def\percentIndexErrorInNonExecutable{1.3\%\xspace}
\def\percentIndexErrorInTotal{1\%\xspace}
\def\totalRuntimeError{388\xspace}
\def\percentRuntimeErrorInNonExecutable{1.1\%\xspace}
\def\percentRuntimeErrorInTotal{0.9\%\xspace}
\def\finalFullyRestored{1,463\xspace}
\def\percentFinalFullyRestored{3.4\%\xspace}
\def\finalPartiallyRestored{3,480\xspace}
\def\percentFinalPartiallyRestored{8.2\%\xspace}
\def\totalFinalRestored{4,943\xspace}
\def\percentTotalFinalRestored{11.6\%\xspace}

%% file: sections/abstract.tex
\begin{abstract}
Computational notebooks are the de facto platform for exploratory data science, offering an interactive programming environment where users can create, modify, and execute code cells in any sequence. However, this flexibility often introduces code quality issues, with prior studies showing that approximately 76\% of public notebooks are non-executable, raising significant concerns about reusability. We argue that the traditional notion of executability—requiring a notebook to run {\em fully} and {\em without error}—is overly rigid, misclassifying many notebooks and overestimating their non-executability.

This paper investigates pathological executability issues in public notebooks under varying {\em notions} and {\em degrees} of executability. Notebooks, by construction, are incrementally and interactively executed, where each cell execution advances logic toward the notebook's goal. Even partially improving executability can improve code comprehension and offer a pathway for dynamic analyses. With this insight, we first categorize notebooks into potentially restorable and pathological non-executable notebooks and then measure how removing misconfiguration and superficial execution issues in notebooks can improve their executability (i.e., additional cells executed without error). For instance, we use a Large Language Model (LLM) to generate synthetic input data to restore non-executable notebooks with ``FileNotFound" errors. In a dataset of \totalNotebooksInDataset popular public notebooks, containing \totalNonExecutable non-executable notebooks, only \percentPathological are truly pathologically non-executable. For restorable notebooks, LLM-based methods fully restore \percentFullyRestored of previously non-executable notebooks. Among the partially restored, it improves the notebooks' executability by \averagePercentModuleNotFoundRestoredIncrease and \avgIncreaseAfterFileFixed by installing the correct modules and generating synthetic data. These findings challenge prior assumptions, suggesting that notebooks have higher executability than previously reported, many of which offer valuable partial execution, and that their executability should be evaluated within the interactive notebook paradigm rather than through traditional software executability standards.
\end{abstract}

%% file: sections/s1_intro.tex
\section{Introduction}
\label{sec:intro}

Computational notebooks are the preferred tool for data analytics due to their portability and interactive capabilities. They enable easy sharing across various contexts and are often made publicly accessible, fostering reuse and adoption \cite{jupyter2021collaboration, Quaranta2022}. In notebooks, developers write code and explanatory text in cells that can be run individually, displaying results instantly. An unexpected consequence of this interactive programming is non-reproducibility---more than 75\% of the public notebooks are previously reported to be non-executable, let alone reproduce the promised output \cite{Pimentel2019}. Such a low level of executability is alarming as it diminishes the accessibility of public codebases and potentially causes code quality issues during reuse \cite{reuseICPC2022}. Previous investigations have studied the executability and reproducibility issue of notebooks and reasoned about the presence of such issues \cite{Pimentel2019, Pimentel2021, Wang2021, Zhu2021, Head2019, Wang2020}. However, the findings of these studies have several limitations. 

\noindent\textbf{\em Problem.} Previous studies categorize notebooks into (1) non-executable if they do not execute end-to-end in their current form or (2) executable if they execute successfully without any interference. There are two limitations to this categorization. First, a seemingly non-executable notebook may technically be executable but lack the appropriate execution environments (input file, packages, or execution order). Second, the binary notion of executability is overly rigid and stringent. Notebooks, by construction, are executed incrementally and interactively, similar to REPL (read-eval-print loop \cite{REPL}), where each cell execution logically progresses the notebook. 

Even if a notebook is non-executable, its partial execution improves code comprehension during reuse. Furthermore, prior works report inconsistent findings on notebook non-executability (76\%\cite{Pimentel2019}, 72.6\%\cite{Wang2021}, 82.6\%\cite{Wang2020}, and 47\%\cite{Zhu2021}) due to differences in the notebooks datasets analyzed. For instance, despite the large number of notebooks studied by prior work \cite{Pimentel2019}, most are unmanaged trial notebooks, often for personal, experimental projects rather than for reuse. Therefore, such findings do not generalize to popular notebooks (i.e., notebooks actively reused), demanding a fresh perspective on the non-executability of notebooks. We hypothesize that the count of executable notebooks is higher than previously reported, and even {\em pathologically non-executable} notebooks are partially executable, offering valuable code reusability, code comprehension, and code repair opportunities.

\noindent\textbf{\emph{Contributions.}} In this paper, we take a fresh perspective on notebook executability by investigating the reusability of public notebooks. We address the gaps in existing knowledge on notebook executability by making the following contributions. 
First, we introduce new notions and degrees of executability by separating misconfiguration from non-executability. To that end, we define {\em pathologically non-executable} notebooks that suffer from unresolvable fatal syntactic or runtime errors (such as bad indentation or attribute error), and {\em restorable} notebooks that are non-executable due to superficial misconfiguration errors (e.g., missing input, library, or execution order). Restorable notebooks are fully executable in a suitable environment (e.g., the original author's local environment).
Second, we introduce fine-grained metrics of executability that quantify the reuse potential of non-executable notebooks. For non-executable notebooks that are otherwise deemed unusable, we measure their {\em partial executability} based on the proportion of successfully executed cells, an aspect that has never been studied.
%
Third, we conduct this investigation on actively reused and adopted notebooks. We follow the practice of using GitHub stars as a popularity metric \cite{Borges2016, KocKleJoh24} to rank notebooks during data collection. Doing so mitigates any bias from unmanaged, rarely used public notebooks. 
Lastly, we demonstrate the effectiveness of LLMs in addressing misconfiguration issues (e.g., by generating synthetic input data) in restorable notebooks.

We realize the aforementioned contributions by building an automated measurement framework for notebooks. Each notebook is evaluated in a sandboxed environment with all required execution environments as stated in the parent repository. We leverage a combination of static and dynamic error checking to extract the causes of non-executability in notebooks. Static analysis on notebook code cells identifies errors such as structural, syntax, and variable undefined errors. Dynamic analysis identifies runtime errors such as missing libraries, packages, and type errors. This stage returns a categorized list of pathologically non-executable notebooks and restorable notebooks and their respective list of errors. LLMs are improving code generation and understanding \cite{Nejjar, lin2024soen101codegenerationemulating}. To investigate the reusability of restorable notebooks, we utilize open-source LLM Llama-3 for error-driven notebook restoration. For example, error reports indicating missing modules or input files are combined with code cells to prompt the LLM to identify the correct module names for installation or generate synthetic input data, respectively. We iteratively restore and error-check the notebook until it is fully executable or an unfixable error is found. 

\noindent\textbf{\emph{Study Results.}} We collect a dataset of the most popular \totalNotebooksInDataset public notebooks from a stratified sample of \totalRepos GitHub repositories with four or more stars in 2024. Using non-executability criteria from prior work, we found that over \percentNonExecutable of those notebooks are non-executable if executed as is, which is consistent with the findings of prior studies \cite{Pimentel2021, Wang2021}. 
We investigate the top two reasons for non-executability: lack of appropriate input files and missing modules, packages, or libraries. Such notebooks are potentially restorable. Surprisingly, we find that {\bf only \percentPathological of all notebooks are pathologically non-executable} due to unrestorable executability issues, and the rest are either executable or potentially executable given a suitable execution environment, i.e., \percentRestorableInNonExecutable of non-executable notebooks can potentially be restored. This validates our hypothesis that pathologically non-executable notebooks are significantly lower than previously reported in prior work. Possible reasons include a skewed notebooks dataset, a misconfigured experiment environment, and the strict notion of executability. 
Interestingly, our error-driven, LLM-guided fix successfully restores full executability of \totalFinalRestored of the notebooks for which it found the necessary module and synthetic input data. It partially improves the executability of restorable notebooks by \averageIncreasePartiallyRestored cells per notebook. Prior studies \cite{Pimentel2019, Pimentel2021} have considered these notebooks completely non-executable. 
Even pathologically non-executable notebooks are not entirely non-executable. They can be successfully executed up to \averagePercentPartialPathological of cells, on average. These findings show that traditional notions of executability underestimate notebooks' reusability mainly due to misconfigured notebooks, some of which can be fixed automatically, and that most notebooks can still offer numerous code reuse and restoration opportunities. We summarize the contributions of this work below. 

\begin{itemize}
    \item We take a fresh look at the reusability of public notebooks by exploring varying interpretations of executability. Our findings offer insights into the partial executability of seemingly non-executable notebooks and distinguish superficial misconfiguration errors from pathological errors. 
    \item This is the first work to incorporate notebook popularity when assessing executability, contextualizing findings to frequently reused and adopted notebooks. 
    \item We report the feasibility of lightweight restoration strategies in improving notebooks' executability by addressing three of the top executability issues. We also demonstrate that LLMs can be effective in generating synthetic sample input to advance the partial executability of notebooks. 
    \item Executability is the first step toward reproducibility. A significant portion of cells in pathologically non-executable notebooks remains executable, unlocking the potential for advancing code reuse and dynamic analysis. Partial executability can serve as a valuable criterion to assess a notebook’s restoration potential. 
\end{itemize}

\noindent We have made our code and dataset publicly available at \url{https://github.com/renote2024/ReNote2024}.

\begin{figure}[t!] 
    \centerline{\includegraphics[width=\columnwidth]{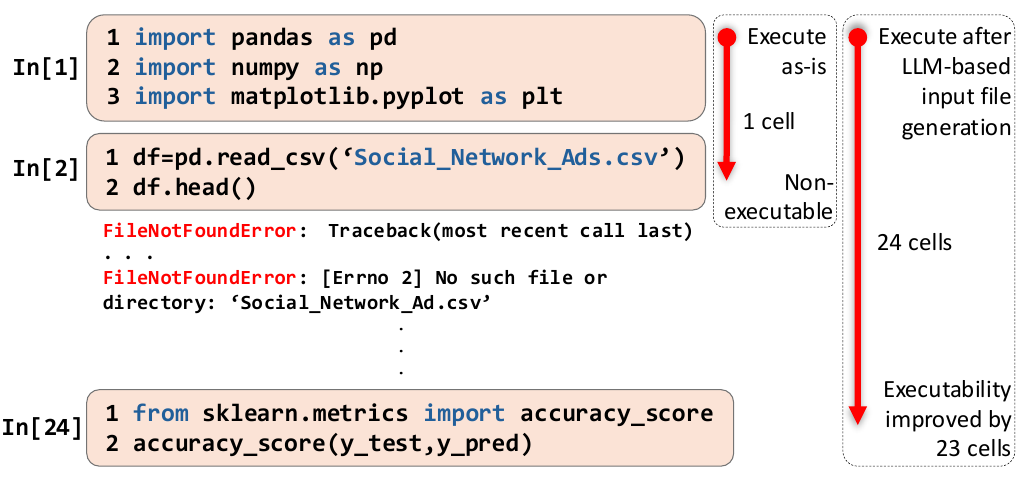}}
    \caption{Notebook \cite{girlscript2024} is initially non-executable due to an undefined variable. Its executability can be fully restored by 23 cells by generating a missing input file for the notebook.}
    \label{fig:cs1-full-exec}
\end{figure}

%% file: sections/s2_motivation.tex
\section{Motivation}
\label{sec:motivation}

\begin{figure*}[ht!] 
\centerline{\includegraphics[width=0.8\textwidth]{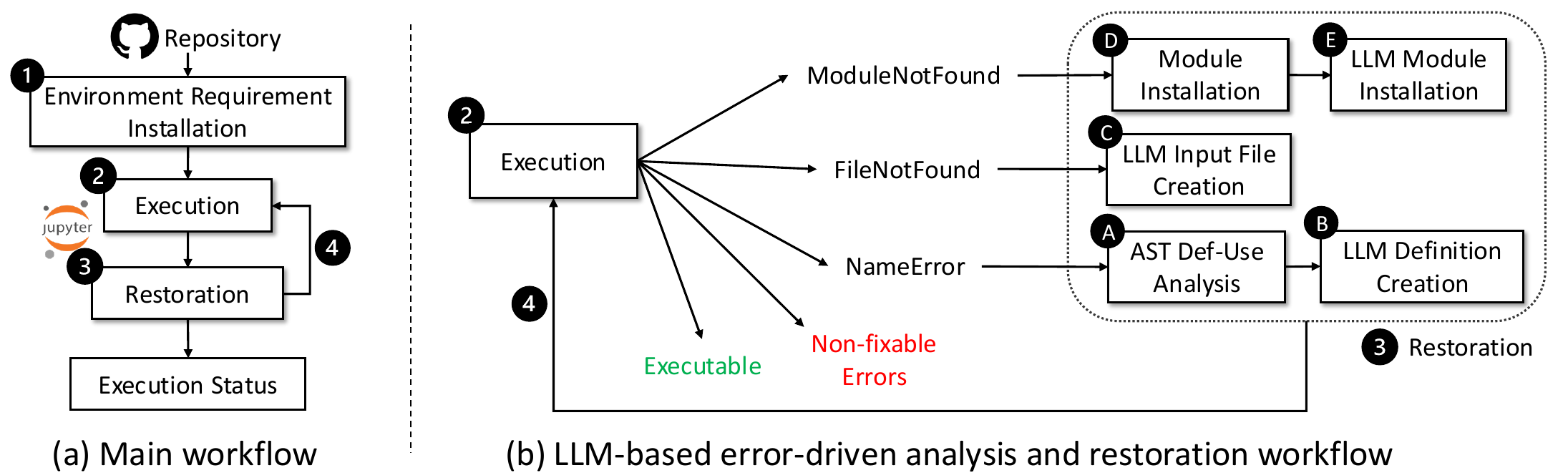}}
    \caption{LLM-based error-driven notebook executability analysis and restoration workflow.}
    \label{fig:main_workflow}
\end{figure*}

This section presents two case studies demonstrating that seemingly non-executable notebooks can potentially be restored with minor reconfiguration, and pathologically non-executable notebooks still offer valuable partially executable code.

\subsection{Case Study 1: Restoring Complete Executability}
\label{case_study_1}

    The notebook {\small{\texttt{random\_forest\_algorithm.ipynb}}} \cite{girlscript2024} implements the random forest classifier algorithm. This notebook is hosted by the GirlScript Foundation's GitHub open-source repository ``Winter of Contributing" with 881 stars. It consists of 24 code cells. When this notebook is executed as is, it results in a ``FileNotFound" error in cell 2, attempting to read input file {\small{\texttt{Social\_Network\_Ads.csv}}}. This input file is accessible in the original author's environment, suggesting that the author intended this notebook to be executable before uploading it to GitHub. However, this input file is neither included in the repository nor available in the adopter's environment. Prior studies \cite{Pimentel2019, Pimentel2021} would classify this notebook as non-executable. This classification is analogous to considering ``{\small{\texttt{javac}}}" non-executable if no input {\small{\texttt{.java}}} file is provided. This notebook aims to demonstrate the logical steps required to create a classifier. Our observation is that for such a notebook, improving executability can help serve the notebook's purpose, which can be done by generating a synthetic input file. To do so, we package the notebook's code, Python error description, and the notebook's documentation into a prompt and query Llama-3 to generate an input file with a correct relative path to the notebook with synthetic content. This synthetically generated input file results in full executability of all 24 cells, i.e., improving the executability of the previously non-executable notebook by over 95\% as illustrated in Figure \ref{fig:cs1-full-exec}. This is a prime example of how non-executability is often over-classified when the input file is not provided but may be available locally to the original author.

\subsection{Case Study 2: Improving Partial Executability}
\label{case_study_2}

    The notebook {\small{\texttt{DinosaurusIsland--Character level language model final-v3.ipynb}}} from GitHub repository ``deep-learning-coursera" \cite{gemaatienza} is demonstrating a deep learning tutorial. It has 129 stars and a total of 14 code cells. When executed in linear order, the notebook first encounters a ``ModuleNotFound" error in cell 1 due to the absence of module {\small{\texttt{utils}}}. We resolve this error by installing the required module. A follow-up execution raises a ``NameError" in cell 6 due to the undefined function {\small{\texttt{softmax()}}}. After reviewing the notebook, we noted that the markdown cell claimed the function was provided, but it is missing. The original notebook shows a valid output, meaning the author successfully executed it. This is a common case where program states are shared between different executions, even if the corresponding code is moved or deleted. Even if the notebook is executed end-to-end before pushing on GitHub, the error would not have surfaced as the function definition is still present in the current Python kernel. However, such program states are not shared with the notebook itself, resulting in a ``NameError" in a new Python Kernel. 
    
    Our analysis indicates that defining the undefined name with an appropriate implementation could significantly improve the notebook's executability. Therefore, we prompt Llama-3 with the error report and the code cells to generate a definition for this function and insert a new code cell containing the definition right before its usage. This LLM-generated code cell requires a module import for {\small{\texttt{tensorflow}}}. Again, we were able to install it in the environment. These simple addition and module installations improve the original notebook's executability by 7 cells until it encounters ``AttributeError" at cell 8. This is a tutorial notebook, and the first half illustrates valuable deep-learning knowledge in sequence models. Thus, partial executability offers added value.

%% file: sections/s3_research_questions.tex
\section{Research Questions}
\label{research questions}

Under the traditional definition of executability, a notebook is considered \textbf{executable} if it does not trigger any error throughout its complete top-down execution. A \textbf{non-executable} notebook, on the other hand, fails to execute due to an error or exception. In this paper, we relax this notion of executability by dividing it further into {\bf pathologically non-executable}, {\bf non-executable but restorable}, and {\bf executable}. When a notebook's executability is hindered by unresolvable errors, e.g., syntax or indentation errors, it is \textbf{pathologically non-executable}. If a notebook is executable in the original author's execution environment but fails to execute in another environment due to issues such as missing input files, execution orders, or execution environments, it is called a  {\bf non-executable but restorable} or {\bf misconfigured} notebook which is not executable but can be fully or partially restored. 

Similar to different notions of executability, we also introduce different degrees of executability. Instead of considering it as a binary metric, we measure executability on a continuous spectrum, defined as a ratio between the number of cells executed until the first error and the total number of cells in a notebook. 100\% executability refers to fully executable notebook, whereas $<$100\% refers to partially executable notebooks. A fully executable notebook does not guarantee reproducibility, i.e., a notebook must produce the same outputs as the original notebook. Reproducibility is beyond the scope of this work and is often not required for data-centric notebooks as they are expected to produce different outputs on different data. We explore the following research questions:

\begin{itemize}
    \item \textbf{RQ1:} What are the common causes of non-executability in notebooks?
    \item \textbf{RQ2:} How many non-executable notebooks are pathologically non-executable, and how many can be restored?
    \item \textbf{RQ3:} To what extent can pathologically non-executable notebooks be executed?
    \item \textbf{RQ4:} Can LLM-based restoration strategies enhance notebook executability?
\end{itemize}

%% file: sections/s4_approach.tex
\section{Empirical Analysis Methodology}
\label{sec:approach}

Figure \ref{fig:main_workflow} (a) illustrates the intermediate stages of our analysis process. Starting with a repository, we retrieve all notebooks and conduct a lightweight static error check to identify issues that may impede notebook execution. Next, we automatically analyze the repository to extract any provided environment requirement files, enabling us to configure the appropriate environment for notebook execution. During execution, we log the first-encountered error and incrementally apply targeted restoration strategies based on the error type, measuring executability improvements at each stage.

\subsection{Analysis Approach}
\label{analysis_approach}

    \subsubsection{Initial Error Checking}
        For a notebook to be executable, it first needs to be compilable. As a first step, we look for compilation errors by applying a lightweight error checking on all notebooks within each repository to uncover any compiling issues, such as syntax or indentation errors. This preliminary filtering step assesses executability without running the code, allowing us to exclude notebooks that require extensive code-level modifications for restoration and are thus deemed pathologically non-executable. To access the content of the notebooks, we use the {\small{\texttt{nbformat}}} module. While performing static analysis of the notebook's code, we detect and omit notebooks from further investigation that (1) cannot be read due to corruption and encoding problems, (2) contain no code cells (with or without markdown cells), or all code cells do not hold any content; and (3) written in programming languages other than Python 3 (such as Python 2, Julia, R, or C).

    \subsubsection{Dynamic Error Checking} 
        \label{dynamic-error-checking}
        In step \ding{203} of Figure \ref{fig:main_workflow} (b), we perform dynamic error checking to detect and categorize potential runtime errors. We use Papermill \cite{papermill}, which allows us to execute the notebook and detect and report any errors encountered during execution. We use Python 3 as the execution kernel and document the first error that halts the execution. We categorize the execution status of each notebook into the following groups:
    
        \begin{enumerate}
            \item \textbf{Executable:} The notebook is fully executable without encountering errors.
            \item \textbf{FileNotFound:} The notebook writes to or reads from a file or directory that is unavailable during execution.
            \item \textbf{ModuleNotFound:} The notebook imports unavailable modules, packages, or libraries.
            \item \textbf{NameError:} The notebook uses a variable, function, or class without defining or importing it.
            \item \textbf{Others/Non-Fixable:} The notebook encounters errors not covered by the above categories. We also identify and categorize these errors accordingly.
        \end{enumerate}

        \begin{figure}[t!] 
            \includegraphics[width=\columnwidth]{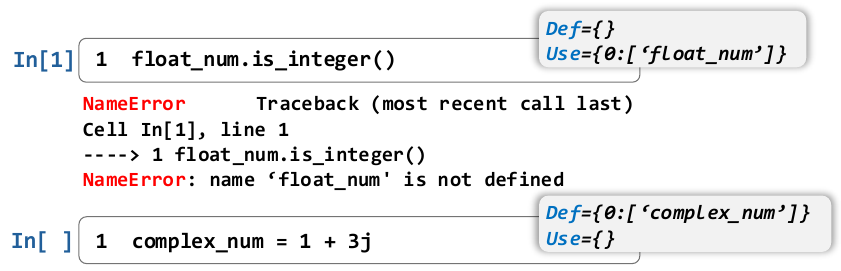}
            \caption{Def-use lists for the first two cells in notebook \cite{pytopia}.}
            \label{fig:def-use}
        \end{figure}

        \noindent The categories reflect a natural progression of notebook adoptions and fixes, as shown in Figure \ref{fig:main_workflow}. 
        If a notebook is found fully executable after this execution process, we mark it as executable. If it is non-executable due to missing input files, required modules, or undefined names, we iteratively apply restoration strategies (Figure \ref{fig:main_workflow} (b)) and perform dynamic error checking to remove as many executability issues as possible.

        For notebooks with ``NameError," we perform a static {\em def-use} analysis to localize the root cause behind undefined names. Like traditional {\em def-use} analysis, we build a specialized AST visitor that tracks variable definition and usage with precise scope awareness within each cell. Figure \ref{fig:def-use} shows how \textit{def} and \textit{use} sets look for two notebook cells. Note that while maintaining the definition set, we consider variable binding, access, and scoping rules for Python to avoid any false positives and false negatives during the analysis. Our definition set extends beyond variables to include imports, functions, classes, and control flow elements. 
        
        Next, we use {\em def-use} sets to isolate the location of undefined names in the notebook and search nearby cells to locate the definition. Finally, the framework returns any detected definition of the undefined variable and its cell location. If the variable's definition appears later in the notebook than its initial use, we categorize the notebook as {\em defined after use} and return the cell number where the definition is found. If no definition exists throughout the notebook, we classify it as {\em undefined}.

\begin{figure*}[t!] 
        \centerline{\includegraphics[width=1\textwidth]{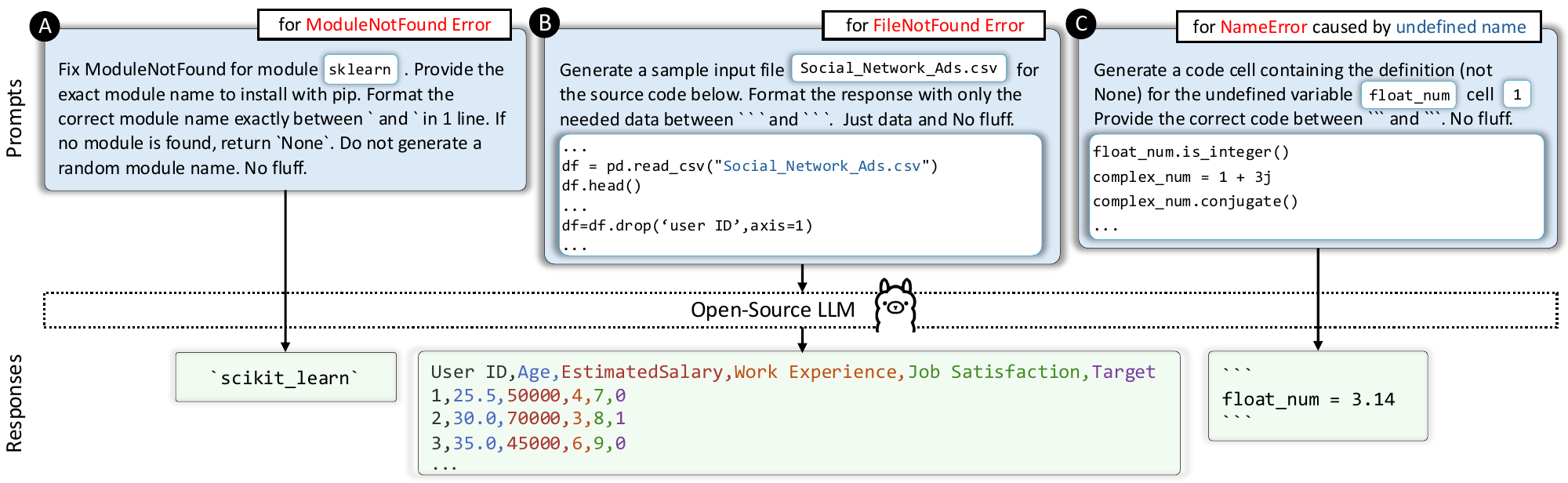}}
        \caption{Examples of prompts for LLM and responses for different error types. }
        \label{fig:LLM-prompts}
\end{figure*}

    \subsection{LLM-Based Error-Driven Restoration}
        This section explains the lightweight use of LLMs to synthesize the execution environments, enabling improved measurements of notebook restorability. This addresses undefined name issues from incorrect execution orders or the absence of definitions, generates synthetic yet syntactically valid input, and installs external modules. 

        \subsubsection{ModuleNotFound Error}
            Most of the repositories do not provide environment requirement information, and if they do, those modules are often outdated \cite{Zhu2021, Wang2021}. Notebooks with ``ModuleNotFound" errors are restorable since they were originally executable but failed to execute in a new environment. We attempt to re-establish the execution environment for the notebooks by inspecting the REQUIREMENTS/INSTALL instructions and installing the required dependencies in the execution environment before executing any notebooks. During execution, if a notebook returns a ``ModuleNotFound" error, we extract the missing module name from the error message and construct a terminal command {\small{\texttt{pip install <missing module>}}} (\ding{204}-D of Figure \ref{fig:main_workflow}). If the installation fails due to an incorrect or deprecated module name, we use LLM to obtain the correct and updated name for the corresponding missing module (\ding{204}-E of Figure \ref{fig:main_workflow} and LLM prompt in Figure \ref{fig:LLM-prompts}-A) and retry the installation.

        \subsubsection{FileNotFound Error}
            ``FileNotFound" errors often occur when notebooks attempt to access and read unavailable input files or directories. We use LLM to generate synthetic input data tailored to the notebook's semantics, attempting to resolve ``FileNotFound" errors while avoiding additional runtime issues. Our insight is that the code cells in notebooks contain sufficient information for LLMs to generate syntactically correct input data needed for execution. Since our goal is executability rather than reproducibility, syntactically correct synthetic data is sufficient to address the ``FileNotFound" error and restore notebook executability.
            Concise contexts improve the LLM's performance \cite{Ramlochan2024}. Thus, we provide contexts of the notebook error, such as the missing file path and type, which guides LLM on the file's syntax, correct data type, and content. Instead of the {\small{\texttt{.ipynb}}} file, we provide the source code of all cells in the prompt. This conversion from {\small{\texttt{.ipynb}}} to Python also removes noise that may misguide or overwhelm the LLMs. Figure \ref{fig:LLM-prompts}-B shows this prompt and the response by the LLM for the example mentioned in \ref{case_study_1}. 

        \subsubsection{NameError}
            For a non-executable notebook due to undefined names, we extract the name and the specific cell in which the error occurs and the cell where the name definition is located or indicate if it cannot be found.  We then prompt LLM to address this issue as illustrated in \ding{204}-A of Figure \ref{fig:main_workflow} (b). The prompt also includes the code cells and the error type (see Figure \ref{fig:LLM-prompts}-C). The LLM's response is used to rewrite the notebook, which is analyzed again for additional errors. However, the changes introduced by LLMs may change the notebook's semantics. Since test cases are almost non-existent in notebooks (only 1.54\% notebooks have tests \cite{Pimentel2021}), it is challenging to verify semantic accuracy. Even if test cases are available, we must first make the notebook executable to enable testing and detect any semantic inconsistencies introduced by LLM-based modifications. Our position is that addressing executability is a prerequisite for reproducibility, which requires cell executability even if it causes semantic changes.

%% file: sections/s5_results.tex
\section{Empirical Results}
The following sections present the empirical findings of the notebook executability analysis.

\input{figures/total_star_vs_count} 

    \noindent{\textbf{Dataset.}} We focus on two criteria to collect notebooks to investigate our research questions. First, the notebooks should be popular, actively shared, and reused. Second, the notebooks should be the most recent version at the time of the notebook dataset collection, as the computational notebook ecosystem has changed in the last few years with the advent of AI and new notebook tooling. We exclude irrelevant files that are either unreadable, written in non-Python languages, or authored in Python 2, which has been deprecated since 2020. Based on this, we use GitHub API to search for Jupyter notebooks with parameters, including language and star range. 
    GitHub stars are commonly used to indicate popularity \cite{Borges2016}.  We then take a stratified sample of up to 1000 repositories based on GitHub star "tiers" (e.g., $\geq$1000, 500-999, 300-499, and so on) with logarithmic tier sizes to balance the repository count per tier due to long-tailed distribution of repositories against stars. This leads to approximately 318,000 notebooks. Due to prohibitive compute costs from LLMs, we take a 13\%  sample of total notebooks,  resulting in \totalNotebooksInDataset notebooks from \totalRepos repositories.    Figure \ref{fig:notebook_count_vs_star} shows the distribution of sampled notebooks based on stars.

\noindent{\textbf{Analysis Environment.}} We establish a separate virtual environment for each repository. Sandboxing the execution environments eliminates the risk of mismatched versions of Python modules, side effects from executing the notebooks, and conflicts stemming from shared dependencies. We use Python version 3 as the execution kernel. Before analyzing a notebook, we locate the requirements file (if available) listing necessary modules or packages and install them in the execution environment (in \ding{202} of Figure \ref{fig:main_workflow}). 
For the feasibility and scalability of this study, we also set an execution timeout of 5 minutes.
We use a state-of-the-art open-source LLM called Llama-3 \cite{llama3, touvron2023llama}.


  

\subsection{RQ1: Causes of Non-Executability}

        \begin{table}[t]
    	\centering
    	\caption{Top 10 Common Errors in Notebooks.}
        \begin{tabular}{|m{3cm}|>{\centering\arraybackslash}m{1.3cm}|>{\centering\arraybackslash}m{1.6cm}|>{\centering\arraybackslash}m{1cm}|}
    		\toprule
    		\textbf{Error Type}       & \textbf{\# of \hspace{10pt}Notebooks}  & \textbf{\% w.r.t Non-Executable}                    & \textbf{\% w.r.t Dataset}              \\
    		\midrule
    		ModuleNotFound Error      & \totalModuleNotFound            & \percentModuleNotFoundInNonExecutable      & \percentModuleNotFoundInTotal \\ 
    		FileNotFound Error        & \totalFileNotFound              & \percentFileNotFoundInNonExecutable        & \percentFileNotFoundInTotal   \\ 
    		AttributeError            & \totalAttributeError            & \percentAttributeErrorInNonExecutable      & \percentAttributeErrorInTotal \\ 
    		ImportError               & \totalImportError               & \percentImportErrorInNonExecutable         & \percentImportErrorInTotal    \\ 
    		ValueError                & \totalValueError                & \percentValueErrorInNonExecutable          & \percentValueErrorInTotal     \\
            TypeError                 & \totalTypeError & \percentTypeErrorInNonExecutable          & \percentTypeErrorInTotal \\
            KeyError                  & \totalKeyError                  & \percentKeyErrorInNonExecutable            & \percentKeyErrorInTotal       \\
            StdinNotImplementedError  & \totalUserInputs                & \percentUserInputsInNonExecutable          & \percentUserInputsInTotal     \\
            IndexError                & \totalIndexError                & \percentIndexErrorInNonExecutable          & \percentIndexErrorInTotal    \\ 
            NameError                 & \totalNameError                 & \percentNameErrorInNonExecutable           & \percentNameErrorInTotal      \\
    	   \bottomrule
	   \end{tabular}
	\label{common_errors}
\end{table}

    Our first goal is to identify the non-executable notebooks and the reason behind their non-executability. 
    We first set up the execution environment by incorporating the installation instructions in the parent repository. This attempt at executing the notebooks results in \totalExecutable (\percentExecutable) executable notebooks, and \totalNonExecutable (\percentNonExecutable) non-executable notebooks. 
    There are two reasons why non-executability is still more than previously reported 76\% \cite{Pimentel2019}. First, repositories often do not provide complete and detailed requirements files. 
    Later, we show that even after following the requirements file, we still encounter environment misconfiguration errors in nearly \percentModuleNotFoundInTotal of the notebooks.
    Second, prior work studied a dataset of notebooks containing a vast amount of obsolete, simple, and unmanaged notebooks; these are experimental notebooks used for dabbling with notebook environments. Such notebooks may not depend on external packages and libraries, which lowers the overall non-executability. Low popularity notebooks tend to be less dependent on external packages and libraries and are often shorter, reducing the chances of ``FileNotFound" and ``ModuleNotFound" errors but exhibiting higher ``NameError" rates, as shown in \cite{Pimentel2019}. We hypothesize that the executability of the low-quality notebooks will be lower once environment configuration errors are addressed. 

     Next, we investigate and categorize why \percentNonExecutable notebooks cannot execute and find the top reasons that hinder their execution. Table \ref{common_errors} presents the top 10 issues encountered in the notebooks. Among them, ModuleNotFound and FileNotFound errors are the most frequent, making up \percentModuleNotFoundInTotal and \percentFileNotFoundInTotal of the entire dataset, respectively. Other common exceptions include ``AttributeError" (\percentAttributeErrorInTotal), ``ValueError" (\percentValueErrorInTotal), ``TypeError" (\percentTypeErrorInTotal), and ``StdinNotImplementedError" (\percentUserInputsInTotal).
    
    

    \subsubsection{Execution Environment}
        In our analysis, \totalModuleNotFound notebooks encountered ``ModuleNotFound" errors during execution when executed in a fresh environment. This accounts for \percentModuleNotFoundInNonExecutable of the total non-executable notebooks. These errors arise when the required modules or packages are not available in the package manager (e.g., {\small{\texttt {pip}}}) or cannot be located within the execution environment in the case of custom packages and libraries. Such mismatches highlight the importance of ensuring compatibility and consistency across execution environments to facilitate seamless notebook execution across different computing setups. Prior work \cite{Pimentel2019} reports that 20.7\% of notebooks failed due to ``ImportError," whereas our analysis finds that only \percentIndexErrorInTotal of notebooks in our dataset encounter this exception. ImportError can occur for various reasons, such as issues within the module or missing dependencies, while the ``ModuleNotFound" error specifically occurs when Python cannot locate the module at all. Introduced in Python 3.6, the ``ModuleNotFound" error clarifies this distinction, making it easier for developers to handle missing modules.

    \subsubsection{External Input Data}
        We find \percentFileNotFoundInNonExecutable of the total non-executable notebooks affected by the ``FileNotFound" error. Additionally, when the notebook requires user-typed inputs, notebook execution will raise a ``StdinNotImplementedError." We do not attempt to restore the executability of those notebooks, as they necessitate dynamic interactions, which are not feasible in an automated execution environment. We detect a total of \percentUserInputsInNonExecutable initially non-executable notebooks that required user inputs during their execution. 
        

        
    \subsubsection{Missing/Disordered Variable Definitions}

        \input{figures/NameError_vs_Stars_Percent_Table} 
        
        ``NameError" in a notebook can arise due to the use of a variable, library, function, or class name that was not previously defined. Among the \totalNonExecutable notebooks that are non-executable, we identify \totalNameError notebooks containing at least one instance of this error, accounting for \percentNameErrorInNonExecutable (\percentNameErrorInTotal of total), as shown in Table \ref{common_errors}. We further investigate these notebooks to explore the relationship between GitHub stars and ``NameError" occurrences. Table \ref{tab:nameerror_star_range} shows the normalized number of notebooks affected by ``NameError" in different star ranges. Our findings indicate that ``NameError" is less prevalent in more popular notebooks, with its occurrence decreasing as the star increases. While ``NameError" affects only a small portion of our dataset, prior work \cite{Pimentel2019} ranks it among the top two errors, reporting it in 14.53\% of notebooks. We observe that ``NameError" is predominantly found in low popularity notebooks, which are less actively reused and frequently maintained, allowing such errors to persist. This provides concrete evidence that ``NameError" tends to exist in less frequently used notebooks. Therefore, a higher ``NameError" rate reported in \cite{Pimentel2019} suggests that {\em the dataset of notebooks studied in earlier work includes a significant number of low-popularity and rarely managed and reused notebooks.} For example, the notebook \cite{JadFatTail}, with only four stars, resembles a trial notebook that defines numerous functions without any documentation, ultimately leading to an undefined function error.

    \subsubsection{Experimentation Notebooks}
        Notebooks are often used as programming playgrounds, and users who are new to programming often use them to dabble with programming. Notebooks resulting from such exercises are not managed and not expected to be shared and, thus, by definition, are incomplete. Similar to these notebooks, we find that \percentTimeout of notebooks in our dataset contains an infinite while loop to demonstrate concepts such as infinite loops, large-scale training campaigns, or I/O operations over the network. Executing such notebooks may result in ``TimeoutError" issues. Notebooks with ``TimeoutErrors" are not analyzable; thus, we exclude those from Table \ref{common_errors}. 

\begin{tcolorbox}[left=0mm, right=0mm, top=0mm, bottom=0mm]
\textbf{Takeaway RQ1.} Our analysis shows that the primary causes of non-executable notebooks are missing dependencies and external data. High-quality notebooks have fewer fundamental errors like ``NameError," indicating better code maintenance and frequent usage. 
\end{tcolorbox}

\subsection{RQ2: Pathological Non-Executability} 
\begin{figure}[t!] 
	\centerline{\includegraphics[width=\columnwidth]{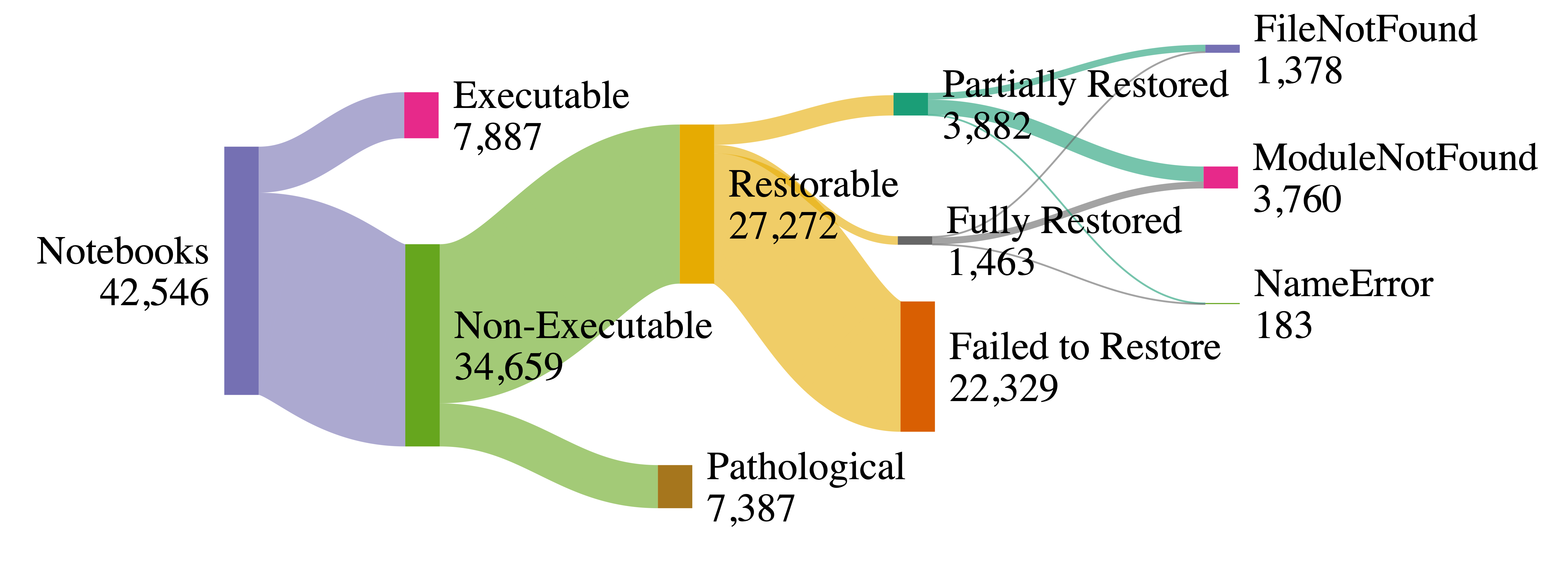}}
	\caption{Summary of our investigation results. This shows different notions of executability in computational notebooks.}
	\label{fig:overall_results}
\end{figure}
 
Resolving errors like compilation errors, syntax errors, indentation errors, and runtime errors like ``OSError'' and ``HTTPError'' require significant code refactoring to align with the intended semantics. Without the knowledge of such semantics, the executability of such notebooks cannot be correctly restored. Notebooks requiring extensive intervention to restore executability are categorized as {\em pathologically non-executable}. Notebooks with errors that can be resolved by configuring the execution environment are classified as non-executable but {\em restorable}. For instance, as in Case Study 1, a missing input data file may trigger a runtime error, rendering the notebook non-executable. However, providing the correct execution environment and data file enables the notebook to run successfully without altering its semantics.

We identify \percentExecutable (\totalExecutable\hspace{1pt}/ \totalNotebooksInDataset) notebooks initially executable, indicating end-to-end execution without errors. Among the rest, shown in Figure \ref{fig:overall_results}, we find \totalPathological (\percentPathological) pathologically non-executable notebooks, distinguished by errors that significantly impede their execution and require additional specification about the intended goal of the notebook to recover. \totalRestorable (\percentRestorableInNonExecutable) non-executable notebooks fall into the restorable category, suggesting errors that could be remedied through appropriate restoration strategies. 

\begin{tcolorbox} [left=0mm, right=0mm, top=0mm, bottom=0mm]
    \textbf{Takeaway RQ2.} Surprisingly, only \percentPathological of all notebooks are pathologically non-executable; the rest are either executable or potentially executable given a suitable execution environment. This validates our hypothesis that pathologically non-executable notebooks are significantly lower than previously found. 
\end{tcolorbox}

\subsection{RQ3: Degree of Executability of Pathologically Non-executable Notebooks}

\input{figures/pathological} 

Notebooks are developed and executed incrementally and interactively, unlike traditional software, which is built and run atomically. Even if a notebook is pathologically non-executable as a whole, it may still contain valuable code snippets that are executable. This introduces a fine-grained notion of executability, where errors may occur only in cells toward the later stages of the notebook. Our position is that these executable cells provide valuable code fragments that can be used for dynamic analysis and supporting tasks such as code reuse, comprehension, repair, and model training. With this new, fine-grained executability, we measure the number of cells a pathologically non-executable notebook successfully executes. We then divide this number by the total number of cells in the notebooks to measure the degree of executability (partial execution) in pathologically non-executable notebooks.

Our analysis reveals that, on average, \averagePercentPartialPathological of the cells in pathologically non-executable notebooks are executable. For example, although the notebook \cite{jarodHAN} encounters ``ValueError" during linear execution, it is still partially executable up to 10 cells or in 76.9\% of total cells. Figure \ref{fig:path-non-exec-results} shows the frequency of pathologically non-executable notebooks with different levels of partial executability. The Y-axis represents the number of notebooks, and the X-axis represents the percentage of cells successfully executed in those notebooks. For instance, we observe that nearly \percentPathologicalOverFiftyPercent of the pathologically non-executable notebooks (\totalPathologicalOverFiftyPercent) successfully execute more than half of the cells ($>$50\%). On the other hand, \totalPathologicalZeroPercent (\percentPathologicalZeroPercent) notebooks have 0\% partial executability due to the pathological error in the first cell. Partial execution in nearly three-quarters of pathological non-executable notebooks demonstrates that even non-executable notebooks can be partially adopted, and dynamic analysis techniques can be applied to improve their executability and reproducibility further.

\begin{tcolorbox} [left=0mm, right=0mm, top=0mm, bottom=0mm]

    \textbf{Takeaway RQ3.} Even pathologically non-executable notebooks contain, on average, \averagePercentPartialPathological executable cells. 
    This demonstrates that partially executable notebooks hold valuable code fragments for downstream restoration steps.
\end{tcolorbox}

\subsection{RQ4: LLM-Based Error-Driven Restoration}
\label{sec:results}
    
    After executing notebooks (Section \ref{dynamic-error-checking}), we categorize \totalNonExecutable non-executable into two main categories: \totalPathological pathologically non-executable notebooks and \totalRestorable restorable notebooks (see Figure \ref{fig:overall_results}). They account for \percentPathological and \percentRestorableInNonExecutable of total non-executable notebooks, respectively.
    This section analyzes the executability of restorable notebooks when restored using lightweight LLM-based error-driven strategies.

    \subsubsection{Addressing ModuleNotFound}   
        Our study reveals that \totalModuleNotFound notebooks encountered ``ModuleNotFound" errors out of \totalRestorable notebooks, accounting for \percentModuleNotFoundInRestorable of the restorable notebooks. We use an error-driven, code-aware prompt to request LLM to infer the missing module name, and then we package the missing model name to install the correct environment for each notebook. During this restoration process, we discover that \totalInvalidModuleNotFound (\percentInvalidModuleNotFoundInAllModuleNotFound) encounter ``ModuleNotFound" errors due to deprecated modules, libraries, and packages \cite{Wang2021, Zhu2021}, changes in installation procedures, or missing custom packages. Old versions of certain packages are sometimes removed from {\small{\texttt{pip}}} repositories due to security vulnerabilities, licensing issues, or to encourage the use of newer, stable releases. This removal makes these versions unavailable, preventing us from resolving dependencies. 
        
        Among the restorable notebook that imports a valid public module, we successfully installed the required modules into the execution environment for \totalModuleNotFoundRestored (resulting in \percentModuleNotFoundRestored success rate), while \totalModuleNotFoundNotRestored failed to do so. Module installation failures occur because only \percentRequirementInTotal of notebooks include an environment requirement file, leading to potential dependency conflicts due to unspecified module versions. For the notebooks where modules were successfully installed, the overall notebook executability increased by \averagePercentModuleNotFoundRestoredIncrease, on average. This improvement underscores the importance of adapting to changes in module availability and installation processes to enhance the executability of notebooks.
    
        Figure \ref{fig:module-found-results} shows the number of notebooks and the corresponding improvements in executability after applying this strategy. The Y-axis represents the number of restored notebooks, and the X-axis represents the increase in executable cells. The figure shows that 500 notebooks have $>$90\% improvements in executability. For example, the notebook \cite{Visualize-ML} initially encounters a ``ModuleNotFound" error due to the absence of {\small{\texttt{networkx}}} module in the environment at cell 1. The environment requirement information is not provided in the repository. This approach restores the full executability of that notebook.

    \subsubsection{Addressing FileNotFound} 

        \input{figures/module_not_found} 
        \input{figures/executable_star_vs_count} 
        Most of the notebooks rely on input data files, which, in most cases, are not packaged with the notebook due to privacy concerns about the data or size. We find that \totalFileNotFound popular notebooks encounter ``FileNotFound" errors, accounting for \percentFileNotFoundInRestorable of non-executable but restorable notebooks. To restore the executability of such notebooks or reveal other execution errors, we use LLM to generate synthetic data for the missing input files automatically.
        As mentioned in Section \ref{sec:approach}, we do not expect synthetic data to reproduce the notebook results. Instead, it can help validate the dynamic behavior of a notebook in order to understand and analyze it better. 
        
        Among the notebooks with ``FileNotFound" errors, we successfully generate input files via LLMs for \totalSuccessfulToGenerateFiles (\percentSuccessfulToGenerateFiles) notebooks. The generated inputs are then placed in a directory structure mentioned in the notebook automatically, and the notebook is executed again. In \totalSuccessfulToGenerateFiles notebooks that were rerun with the synthetic input files, the executability of \totalFileNotFoundRestored (\percentFileNotFoundRestoredInSuccessfullyGenerated) notebooks are fully or partially restored.
        
        Our dataset of notebooks performs a wide range of tasks (statistics, ML, AI, and data analysis) across a wide range of domains, which leads to a large variety of input files being processed by these notebooks. Figure \ref{fig:cs1-full-exec} shows a real-world scenario where a notebook is fully restored after generating synthetic input file {\small{\texttt{Social\_Network\_Ads.csv}}}. The complexity of the content of this file (\eg textual data as column names, numerical data as column values, precision in generating binary labels for the {\small{\texttt{Target}}} column, and avoiding alphabetic words in numerical columns) is shown in Figure \ref{fig:LLM-prompts}. Thus, \percentFileNotFoundRestoredInSuccessfullyGenerated increase in notebook execution with synthetic input files is significant and a promising direction to improve notebook executability (Section \ref{sec:discussion}).

    Figure \ref{fig:llm-input-results} shows the enhancement in executability achieved in non-executable notebooks with LLM-based input generation. The X-axis represents the percentage of the increase of successfully executed cells. The Y-axis is the corresponding number of notebooks restored. For instance, 51 notebooks see more than 90\% increase in executability, whereas \totalFileNotFoundLessFivePercent show less than 5\%. Among notebooks for which we were able to generate the synthetic input files, we partially restore \percentFileNotFoundPartiallyRestoredInSuccessfullyGenerated of the notebooks and fully recover \percentFileNotFoundFullyRestoredInSuccessfullyGenerated of them using the LLM-based input generation. On average, we observe an incremental improvement of \avgIncreaseAfterFileFixed in the number of executable cells in a notebook after employing the LLM-based restoration.

\subsubsection{Addressing NameError}
   We find only \totalNameError notebook encountering ``NameError," among which only \totalDefinedAfter had define-after-use issues. Using LLM to address this issue in a few notebooks will not offer statistically significant insights. Instead, we use LLM to assist in generating proper definitions for the undefined variables or names. Among \totalNameError, we enhance the executability of \totalNameErrorFixed notebooks (\percentNameErrorFixedInAllNameErrors) by applying this approach. This comprises \percentNameErrorFixedFullInAllNameErrors and \percentNameErrorFixedPartialInAllNameErrors notebooks fully and partially restored in the same order. Case Study 2 in Section \ref{case_study_2} provides an example of a notebook facing ``NameError" due to an undefined function, which prevents full execution. We use LLM to identify a definition for this function under the {\small{\texttt{tensorflow}}} module, improving executability by 50\%. This increase is an indication of LLM's potential for restoring functionality in notebooks with missing definitions.


\begin{tcolorbox} [left=0mm, right=0mm, top=0mm, bottom=0mm]
    \textbf{Takeaway RQ4.} Utilizing LLM-based error-driven restoration strategies can potentially enhance the executability of restorable notebooks. For notebooks encountering errors such as ``ModuleNotFound," ``FileNotFound," and ``NameError," we achieve improvements of \averagePercentModuleNotFoundRestoredIncrease, \avgIncreaseAfterFileFixed, and \avgIncreaseAfterNameFixed in executability for each respective exception.
\end{tcolorbox}

\noindent Finally, we summarize the impact of restoration in Figure \ref{fig:executable_vs_star}, which shows the increase in the number of notebooks that achieved full executability before and after all restorations have been applied.


%% file: figures/total_star_vs_count.tex
\begin{figure}[t]
    \centering
    \begin{tikzpicture}
        \begin{axis}[
            width=1\columnwidth, 
            height=0.40\columnwidth, 
            ybar,
            symbolic x coords={$\geq$1000, 500-999, 300-499, 200-299,150-199,125-149,100-124,90-99,80-89,70-79,60-69,55-59,50-54,45-49,40-44,35-39,30-34,25-29,20-24,15-19,10-14,4-9},
            xtick=data,
            ylabel={\# of Notebooks},
            xlabel={GitHub Stars},
            ymin=0,
            ymax=14000, 
            ytick={0, 2000, 4000, 6000, 8000, 10000, 12000},
            minor tick num=0, 
            nodes near coords,
            every node near coord/.append style={font=\scriptsize, yshift=-2pt, xshift=-3pt, rotate=75, anchor=west},
            enlarge x limits=0.05,
            bar width=0.17cm,
            xticklabel style={font=\scriptsize, rotate=75},
            yticklabel style={font=\scriptsize}, 
            label style={font=\small}
        ]
        \addplot[
            fill=blue,
            draw=none
        ] table[x=Star,y=Count,col sep=space]{starVScount.csv};

        \end{axis}
    \end{tikzpicture}
    \caption{Distribution of notebooks and their GitHub stars.}
    \label{fig:notebook_count_vs_star}
\end{figure}

%% file: figures/NameError_vs_Stars_Percent_Table.tex


\begin{table}[t]
    \centering
    \resizebox{\columnwidth}{!}{ 
        \begin{tabular}{|l|c|c|c|c|c|}
            \toprule
            \textbf{Star Range} & \textbf{$\geq$1000} & \textbf{500-999} & \textbf{100-499} & \textbf{10-99} & \textbf{4-9} \\
            \midrule
            \textbf{Percentage} & 0.65\% & 0.41\% & 1.11\% & 1.05\% & 2.09\% \\
            \bottomrule
        \end{tabular}
    }
    \caption{Percentage of notebooks with NameError in different star ranges.}
    \label{tab:nameerror_star_range}
\end{table}

%% file: figures/pathological.tex
\begin{figure}
    \centering
    \begin{tikzpicture}
        \begin{axis}[
            width=1\columnwidth, 
            height=0.45\columnwidth, 
            ybar,
            symbolic x coords={$>$0-9, 10-19, 20-29, 30-39, 40-49, 50-59, 60-69, 70-79, 80-89, 90-99},
            xtick=data,
            ylabel={\# of Notebooks},
            xlabel={Executable Cells (\%)},
            ymin=0,
            ymax=1200, 
            ytick={0, 200, 400, 600, 800, 1000},
            minor tick num=0, 
            nodes near coords,
            every node near coord/.append style={font=\scriptsize, yshift=-2pt},
            enlarge x limits=0.1,
            bar width=0.4cm,
            label style={font=\small},
            xticklabel style={font=\scriptsize, rotate=45}, 
            yticklabel style={font=\scriptsize}, 
        ]
        \addplot[
            fill=blue,
            draw=none
        ] table[x=Executable,y=Count,col sep=space]{pathological.csv};
        \end{axis}
    \end{tikzpicture}
    \caption{Executability of pathologically non-executable notebooks in terms of the percentage of successfully executed cells. }
    \label{fig:path-non-exec-results}
\end{figure}
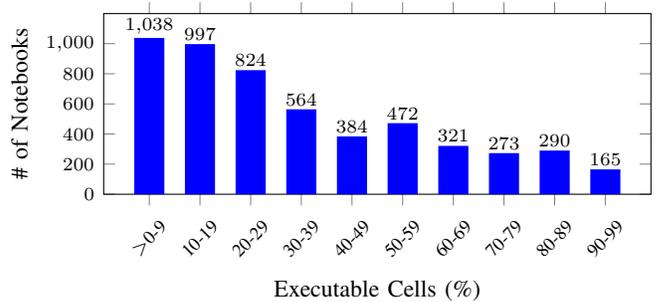

%% file: figures/module_not_found.tex
\begin{figure}[t]
    \centering
    \begin{tikzpicture}
        \begin{axis}[
            width=1\columnwidth, 
            height=0.45\columnwidth, 
            ybar,
            symbolic x coords={$>$0-10, 11-20, 21-30, 31-40, 41-50, 51-60, 61-70, 71-80, 81-90, 91-100},            
            xtick=data,
            ylabel={\# of Notebooks},
            xlabel={Executable Cells (\%)},
            ymin=0,
            ymax=970, 
            ytick={0, 200, 400, 600, 800},
            minor tick num=0, 
            nodes near coords,
            every node near coord/.append style={font=\scriptsize, yshift=-2pt},
            enlarge x limits=0.1,
            bar width=0.4cm,
            label style={font=\small},
            xlabel style={yshift=2pt},
            xticklabel style={font=\scriptsize, rotate=45}, 
            yticklabel style={font=\scriptsize}, 
        ]
        \addplot[
            fill=red,
            draw=none
        ] table[x=Executable,y=Count,col sep=space]{modulenotfound.csv};

        \end{axis}
    \end{tikzpicture}
    \caption{Improvement in executability by addressing ModuleNotFound errors.}
    \label{fig:module-found-results}
\end{figure}
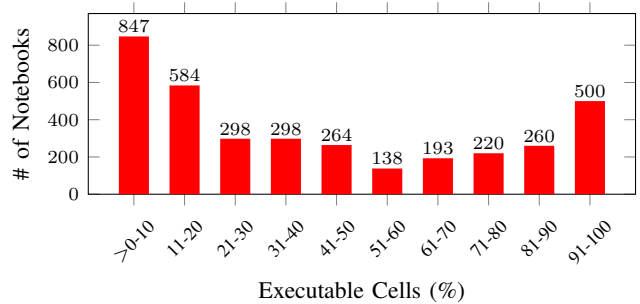

%% file: figures/executable_star_vs_count.tex
\begin{figure*}[!t]
    \centering
    \begin{minipage}{0.32\textwidth}
        \centering
        \begin{tikzpicture}
        \begin{axis}[
            width=\columnwidth, 
            height=0.65\columnwidth, 
            ybar,
            symbolic x coords={$>$0-10, 11-20, 21-30, 31-40, 41-50, 51-60, 61-70, 71-80, 81-90, 91-100},
            xtick=data,
            ylabel={\# of Notebooks},
            xlabel={Executable Cells (\%)},
            ymin=0,
            ymax=450, 
            ytick={0,100,200, 300, 400},
            minor tick num=0, 
            nodes near coords,
            every node near coord/.append style={font=\scriptsize, yshift=-2pt},
            enlarge x limits=0.1,
            bar width=0.25cm,
            label style={font=\small},
            xlabel style={yshift=2pt},
            xticklabel style={font=\scriptsize, rotate=45}, 
            yticklabel style={font=\scriptsize}, 
        ]
        \addplot[
            fill=purple,
            draw=none
        ] table[x=Executable,y=Count,col sep=space]{filenotfound.csv};
        \end{axis}
    \end{tikzpicture}
    \caption{Improvement in executability with synthetic input. 
    }
    \label{fig:llm-input-results}
    \vspace{-5ex}
    \end{minipage}
    \hfill
    \begin{minipage}{0.668\textwidth}
        \centering
        \begin{tikzpicture}
        \begin{axis}[
            width=\columnwidth, 
            height=0.35\columnwidth, 
            ybar,
            symbolic x coords={$\geq$1000, 500-999, 300-499, 200-299,150-199,125-149,100-124,90-99,80-89,70-79,60-69,55-59,50-54,45-49,40-44,35-39,30-34,25-29,20-24,15-19,10-14,4-9},
            xtick=data,
            ylabel={\shortstack{\# of Executable \\ Notebooks}},
            xlabel={GitHub Stars},
            ylabel style={yshift=0pt},
            xlabel style={yshift=5pt},
            ymin=0,
            ymax=3100, 
            ytick={0, 1000, 2000, 3000},
            yticklabels={0,1,2,3}, 
            extra y tick style={tick label style={anchor=south, yshift=5pt}}, 
            extra y ticks={3000}, 
            extra y tick labels={$ .10^3$},
            minor tick num=0, 
            nodes near coords,
            every node near coord/.append style={font=\scriptsize, rotate=90, anchor=west},
            enlarge x limits=0.03,
            bar width=0.12cm,
            xticklabel style={font=\scriptsize, rotate=45}, 
            yticklabel style={font=\scriptsize}, 
            label style={font=\small, yshift=0},
            legend style={
                at={(0.7,0.7)}, 
                anchor=south, 
                font=\scriptsize,
                legend columns=2, 
            }
        ]
      
        \addplot[
            fill=blue,
            draw=none
        ] table[x=Star,y=Count,col sep=space]{InitialstarVScount.csv};
        \addlegendentry{Initial Count};

        \addplot[
            fill=orange,
            draw=none
        ] table[x=Star,y=Count,col sep=space]{FinalstarVScount.csv};
        \addlegendentry{Final Count};
        
        \end{axis}
    \end{tikzpicture}
    \caption{Fully executable notebooks before/after restorations. 
    }
    \label{fig:executable_vs_star}
    \end{minipage}
\end{figure*}

%% file: sections/s6_discussion.tex
\section{Discussion}
\label{sec:discussion}

    \begin{table*}[t!]
    \centering
    \caption{Summary of some related works with ours in computational notebook topic.}
    \begin{tabular}{|m{2cm}|>{\raggedright}m{1.6cm}|>{\centering\arraybackslash}m{0.8cm}|>{\centering\arraybackslash}m{1cm}|>{\centering\arraybackslash}m{1.5cm}|>{\centering\arraybackslash}m{1.3cm}|>{\centering\arraybackslash}m{2.0cm}|>{\raggedright\arraybackslash}m{4.5cm}|}
        \toprule
        
        \textbf{} & \textbf{Purpose} & \textbf{Study Dataset} & \textbf{Module Error} & \textbf{NameError / Cell Order} & \textbf{Input File Error} & \textbf{Non-Executable Found} & \textbf{Dataset Characteristics}\\ 
        \midrule
        
        RELANCER \cite{Zhu2021}                          & Executability        & 4,043             & \cmark    & \xmark    & \xmark    & 47\%      & Collected from Meta Kaggle \cite{kaggle}\\ 
        \hline
        SnifferDog\cite{Wang2021}                        & Executability        & 2,646             & \cmark    & \xmark    & \xmark    & 72.6\%    & Random sample from \cite{Pimentel2019} only those with installable dependency\\ 
        \hline
        Osiris\cite{Wang2020}                            & Reproducibility      & 5,393             & \xmark    & \cmark    & \xmark    & 82.6\%    & Random sample from \cite{Pimentel2019}\\ 
        \hline
        Pimentel et al.\cite{Pimentel2019, Pimentel2021} & Empirical            & $>$1.4M    & N/A       & N/A       & N/A       & 76\%      & Includes high-number of low-starred, rarely reused notebooks from GitHub\\ 
        \hline
        \textbf{This work}                               & Executability        & 42.5K              & \cmark    & \cmark    & \cmark    & \percentPathological & Highly-starred notebooks from GitHub\\ 
        \bottomrule
    \end{tabular}
    \label{related_work_table}
\end{table*}

    In this section, we distill key findings in a {\em FAQ} format, demonstrating the value of accurate executability categorization, the utility of a notebook corpus once marked unusable, and practices that could prevent non-executability or restore executability in future notebooks. 
    

\noindent{\em Why are \percentPathological notebooks pathologically non-executable and \percentPartiallyRestored of restorable notebooks still only partially executable?} 
    Prior work has identified that notebooks have alarmingly low numbers of test cases \cite{Pimentel2019}. Developers releasing their notebooks for public use have very limited testing tools to verify reproducibility and executability. For example, notebooks with undefined variables ``NameError" and ``AttributeError" often go unnoticed due to unintended dependencies on session states saved from prior cell execution, leading to those errors being masked on the developer's machine, similar to Case Study 2 in Section \ref{case_study_2}. In a new environment, such states are unavailable, resulting in these errors. 
    Our findings on pathological errors in notebooks demonstrate the need for static and dynamic analysis tools to catch such errors early in notebook development. 

\noindent{\em How can LLMs further restore notebook executability with a high success rate?} 
    In the first case study (Section \ref{case_study_1}), Llama-3 successfully generated synthetic data for a `.csv' file. However, in another case \cite{tirthajyoti}, it failed to produce a valid `PNG' file due to limitations in multi-modal generation. This is because most LLMs have been trained on similar textual data formats. Therefore, Llama-3 also faces limitations when generating images, audio, video, or zip files. Multi-modal models such as GPT-4o can generate richer and more diverse inputs. 
    This capability can benefit developers who want to share their notebooks without disclosing the input data files. They can leverage LLMs to generate synthetic data (or scripts for synthetic data generation similar to database benchmarks \cite{tpcds}), enhancing executability in remote environments. Additionally, LLM performance tends to degrade when provided with large contexts, which aligns with recent research on LLMs \cite{Leng2024}. Moreover, notebook executability can be enhanced by generating feedback-driven fixes (e.g., multi-shot) for non-executable cells. This includes generating new input files based on error feedback.  


\noindent{\em What use is there for partially executable notebooks?}
    Prolonging notebook execution leads to more cells being executed successfully, which translates into more code surfaces that can be executed and understood. This inherently improves code reuse and enhances the chances of notebooks being fully restored. Furthermore, dynamic analysis techniques are limited to executable code only. By improving notebook executability, we increase the application surface of dynamic analysis tools such as dynamic taint analysis \cite{clause2007dytan}, runtime tracing \cite{meliou2011tracing}, automated debugging \cite{parnin2011automated}, symbolic execution \cite{baldoni2018survey}, and Osiris \cite{Wang2020} that can play vital roles in recovering notebook reproducibility. 
    Partial executability metrics can also provide fine-grained measures of the effectiveness of reproducibility and execution-restoration tools. Additionally, it can serve as valuable feedback when applying code repair techniques in notebooks, such as automated cell reordering \cite{Wang2020}.

\noindent {\em How do the findings of this study improve notebooks?} 
    We explored several design choices in building our measurement framework. We identify gaps in the notebook ecosystem for code review, debugging, test generation, and build tools. For existing public notebooks, which are growing exponentially \cite{Rule2018}, this measurement framework can be extended to improve fine-grained executability, feeding into dynamic analysis tools to enhance reproducibility. 
    
    Our findings, showing that only \totalRequirement (\percentRequirementInTotal) repositories have REQUIREMENTS files, indicate that there is no standardized build and continuous integration mechanism for notebooks. While this may not apply to exploratory, standalone, one-off notebooks, the most popular repositories provide a range of notebooks that must be properly orchestrated to ensure executability. We believe Python-based build tools, such as {\small{\texttt{disutils}}} \cite{distutil} and {\small{\texttt{setuptools}}} \cite{setuptools}, can be extended to support notebook extensions at the browser level to enforce correct build and dependency management for notebooks.

 \noindent {\em Threats to validity.}
    Outdated, textual, and empty notebooks can cause variations in the results of this study. To mitigate that, we exclude notebooks written in Python 2 since support for Python 2 was discontinued in 2020. We also filter empty or instructional notebooks from our dataset. Similarly, focusing purely on GitHub repositories can bias the results towards a specific class of notebooks. Upon investigation, we identified numerous HuggingFace and Kaggle notebooks that are also hosted on GitHub. While there is always room to increase the scale of the analysis, our emphasis on popular, actively reusable notebooks deprioritizes the need to expand the scale. Our static and dynamic analyses rely on the Python AST parser and Python interpreter to capture errors. These tools can potentially miss or misclassify errors, which can impact our categorization. 
    Since most of our runtime errors are captured through dynamic error checking, an earlier fatal runtime error (that we cannot resolve) may hinder our ability to capture ``FileNotFound" or ``ModuleNotFound" errors. Lastly,  we use a timeout of 5 minutes to execute each notebook. 
    This approach may exclude notebooks like the ones requiring expensive machine learning training. However, in our notebook corpus, only \percentTimeout of notebooks encountered timeout errors.

%% file: sections/s7_related_work.tex
\vspace{-0.5ex}
\section{Related Work}
\label{related work}

    With the increasing popularity of notebooks \cite{Kluyver2016}, a growing body of research has focused on understanding the unique characteristics of notebooks \cite{Chattopadhyay2020, Rule2018, Rule2018CellFolding, In2024, gigascience2024} and their diverse applications in various fields \cite{Wang2019, Randles2017, Li2021}. Recent works have also proposed potential enhancements to these tools, aiming to improve notebooks' functionality and usability for data scientists and other users \cite{Chattopadhyay2020, Wang2024, McNutt2023, gigascience2024}. Table \ref{related_work_table} summarizes and compares this work with prior investigations. 

    \subsection{Empirical Studies on Notebook Executability}
        Pimental et al. \cite{Pimentel2019, Pimentel2021} conducted an analysis of 1.4 million notebooks sourced from GitHub, aiming to examine their nature, quality, and reproducibility. They report notebook quality results, covering various characteristics and insights into notebook execution processes, as well as common issues encountered within these notebooks. Similarly, Wang et al. \cite{Wang2019BetterCode} assesses the quality of code present in notebooks, ultimately concluding that notebooks often exhibit suboptimal coding practices, thus highlighting the importance of enhancing notebook code quality. Both studies only view notebook executability as an atomic measure. In particular, they lack a fine-grained, deeper analysis of a large portion of notebooks that are (1) non-executable only on non-author environments due to misconfiguration, thus fully restorable, and (2) offer reusable, valuable code that is partially executable.

    \subsection{Notebook Restoration and Reproducibility Tools}
        Numerous tools are proposed to restore notebook executability and reproducibility. For instance, RELANCER automatically updates deprecated APIs in non-executable notebooks by gathering APIs from GitHub and documentation \cite{Zhu2021}. Similarly, SnifferDog restores the execution environment by providing a collection of APIs from Python packages and libraries to update required packages in notebooks \cite{Wang2021}. These approaches complement our analysis. However, due to the highly complex and heavyweight nature of these techniques, they are infeasible for large-scale notebook executability analysis. 
        Osiris \cite{Wang2020} restores the reproducibility of only fully executable notebooks that also contain output cells. It addresses cell dependencies and generates possible execution orders, similar to our def-use set-based reordering. Osiris demonstrates that without executable notebooks, reproducibility can not feasibly be restored. Similar to our investigation, a transparent, fine-grained view of the state of notebook executability can facilitate further reproducibility studies or tool development.

    \subsection{Notebook Development Assistance Tools}
        Previous research has proposed improving software development practices with extended user interfaces and advanced programming assistant features. Fork It \cite{Weinman2021} is a forking and backtracking extension designed to investigate various alternatives and traverse different states within a single notebook. Similarly, numerous notebook tools \cite{Rule2018CellFolding, Li2024, Zhu2024Facilitating, Wang202Conflict, SuperNOVA} aim to streamline collaborative and exploratory experiments, often incorporating new visualization and conflict resolution for collaborative work in notebooks. To assist in code debugging and cleaning,  Robinson et al. \cite{Robinson2022} examined error identification strategies used for Python notebooks, while Head et al. \cite{Head2019} utilized program slicing to extract relevant code cells producing specific outputs.  
        Adding new features and complicated user interfaces significantly intervenes with rapid, interactive programming. Most of these tools have seen resistance in adoption by both users and popular notebook platform providers. With the emergence of LLMs, recent work has explored using LLMs in assisting notebook development \cite{McNutt2023, Wang2024, Weber2024Computational, grotov2024untangling}. These results resonate with our findings.

%% file: sections/s8_conclusion.tex
\section{Conclusion}
\label{sec:conclusion}

Computational notebooks, widely used for data science and ML/AI tasks, continually suffer from executability issues. Prior studies reporting the non-executability of notebooks rely on a rigid definition of executability, leading to overestimating non-executable notebooks. For the first time, we introduce the notions of partial executability and pathological non-executability for notebooks, contextualizing executability to the notebook's interactive computing paradigm. Our investigation finds that \totalPathological out of \totalNotebooksInDataset notebooks are pathologically non-executable, while \totalRestorable can be restored given suitable execution environments. We leverage LLM-based error-driven restoration techniques to fully restore \percentFullyRestored and partially restore \percentPartiallyRestored of previously non-executable notebooks. These results offer key evidence that notebooks can benefit from LLM-based restoration, and partial executable notebooks are still valuable for broader code reuse practices.

\section*{Acknowledgement} We thank anonymous reviewers for providing valuable and constructive feedback to help improve the quality of this work. This work was supported in part by Amazon - Virginia Tech Initiative in Efficient and Robust Machine Learning, 4-VA, and the National Science Foundation award 2106420. We also thank the Advanced Research Computing Center at Virginia Tech for their support in building and evaluating this work.